\def\year{2023}\relax
\documentclass[letterpaper]{article} 
\usepackage{aaai22}  
\usepackage{times}  
\usepackage{helvet}  
\usepackage{courier}  
\usepackage[hyphens]{url}  
\usepackage{graphicx} 
\usepackage{natbib}  
\usepackage{caption} 
\DeclareCaptionStyle{ruled}{labelfont=normalfont,labelsep=colon,strut=off} 
\usepackage{algorithm}
\usepackage{algorithmic}
\usepackage{newfloat}
\usepackage{listings}
\usepackage{multirow}
\usepackage{amssymb}
\usepackage{pifont}
\usepackage{bibentry}
\newcommand{\cmark}{\ding{51}}
\newcommand{\xmark}{\ding{55}}
\usepackage{array}
\newcolumntype{P}[1]{>{\centering\arraybackslash}p{#1}}

\floatstyle{ruled}
\newfloat{listing}{tb}{lst}{}
\floatname{listing}{Listing}
\title{On Achieving Privacy-Preserving State-of-the-Art Edge Intelligence}
\author{
    Daphnee Chabal\textsuperscript{\rm 1}\thanks{This work was partially supported by the European Union’s Horizon 2020 research and innovation programme under grant 871525 (FogProtect).}\thanks{Presentation of this work was partially supported by the European Union’s project SECURED.}, Dolly Sapra\textsuperscript{\rm 2}, Zolt\'an \'Ad\'am Mann\textsuperscript{\rm 1}
}
\affiliations{
    \textsuperscript{\rm 1}Complex Cyber Infrastructure group, University of Amsterdam\\
    Lab42, Science Park 900, 1012 WX Amsterdam, The Netherlands

\textsuperscript{\rm 2}Parallel Computing Systems group, University of Amsterdam\\
    Lab42, Science Park 900, 1012 WX Amsterdam, The Netherlands
}

\begin{document}

\maketitle

\begin{abstract}
Deep Neural Network (DNN) Inference in Edge Computing, often called Edge Intelligence, requires solutions to insure that sensitive data confidentiality and intellectual property are not revealed in the process. Privacy-preserving Edge Intelligence is only emerging, despite the growing prevalence of Edge Computing as a context of Machine-Learning-as-a-Service. Solutions are yet to be applied, and possibly adapted, to state-of-the-art DNNs. This position paper provides an original assessment of the compatibility of existing techniques for privacy-preserving DNN Inference with the characteristics of an Edge Computing setup, highlighting the appropriateness of secret sharing in this context. We then address the future role of model compression methods in the research towards secret sharing on DNNs with state-of-the-art performance.
\end{abstract}

\section{Introduction}
\noindent Deep Neural Networks (DNNs), the prominent tools used in the field of Artificial Intelligence, are sought after in many sectors of activities to optimize decision-making and improve the quality of services \cite{lin2022survey}. Specifically, the amount of DNN deployments for commercial purposes during a customer's interaction with everyday objects is proliferating \cite{24}.

Privacy-preserving Inference aims to protect the privacy and security of data belonging to the multiple parties involved in Neural Network Inference. 

There is a global rise of smart services offered by internet-connected devices (sometimes called the Internet of Things or IoT), which are increasingly immersed in daily life (e.g., smartwatches, smartphones, personal digital assistants), and recording confidential facts about our lives. The International Data Corporation estimates that in 2025 there will be more than 55 billion IoT devices in the world \cite{daily_iot}, compared to 12.5 billion in 2010 \cite{137}. The data these devices collect at the edge of the edge-cloud computing continuum will, in many use cases, be processed locally, through an emerging decentralized computing paradigm called Edge Computing \cite{yu2017survey, 108, 168, ayed2021fogprotect}.

The privacy risk in processing the data through DNNs is two-fold. On one hand, Inference data is produced by individuals, institutions, businesses, and is held by the devices they own or use, and may be shared with the businesses that make those devices available. The data however needs to be shared in full with parties that facilitate the "intelligence", as is the case for the Machine-Learning-as-a-Service (MLaaS) business model. On the other hand, DNNs are costly for companies to develop. The DNNs architecture, inner parameters, as well as the sensitive features contained in the data used during training are then deemed valuable confidential proprietary data for the companies. The model however still needs to be made available to third parties to generate meaningful (i.e., accurate) Inference outputs. 

In recent years, many techniques have been put forward to solve the predicament of functional-yet-privacy-preserving DNN Inference \cite{2, 220}. Most works however do not consider the global context of secure and private AI deployment for MLaaS, in terms of (1) the characteristics of distributed systems that execute DNN Inference and (2) the computational requirements of actual state-of-the-art DNNs underlying commercial smart services. 

Edge Computing offers several advantages over cloud computing, including reduced latency for better user experience and increased agency over the data’s life cycle, as data is redirected through fewer nodes, is less attainable to unknown third parties, and risks of bottleneck in gateways decrease \cite{133, 134}.
However, a major drawback is that the devices involved, with Inference clients such as IoT objects or sensors, and DNN-holders such as Edge servers or small data centers, have less computational capacity than that offered on demand by cloud platforms \cite{xu2021Edge, 139,ayed2022protecting}. Moreover, methods of privacy-preserving AI are assumed to be applied to systems already equipped with ubiquitous security procedures existing in distributed systems globally (e.g., access control, anomaly detection, encrypted communication) and that themselves add load \cite{aqeel2016security,lachner2021towards}.

Some work \cite{70, 285, yan2019privacy} has been proposed to bring privacy-preserving DNN Inference to de-centralized Edge Computing. However, these methods were evaluated with outdated DNNs that are less computationally complex than state-of-the-art models we see in present-day AI applications. These simpler models have little to no real-world application in commercial MLaaS setups.

The aim of this paper is to present informed recommendations for upcoming research to achieve privacy-preserving DNN Inference in modern and commercially relevant Edge Computing settings. While promising, works emerging in this domain are still isolated efforts. The research avenues we formulate, which we coin here as \textbf{Privacy-Preserving Edge Intelligence}, are at the emerging intersection of two very active research fields, privacy-preserving DNN Inference and DNN Inference in Edge Computing. 

It is important to note that the training phase is out of scope for this paper as training is impractical in Edge Computing settings, especially in the context of MLaaS. In particular, Federated Learning is a promising solution already put forward for computationally-sensitive privacy-preserving training in a collaborative setting and is actively researched \cite{ yin2021comprehensive, mothukuri2021survey, dp_new}, but is not in our scope as we focus on inference. 

\section{Privacy Requirements for Edge Intelligence}
Devices and Edge servers have a high risk of malicious tampering and interventions due to their ease-of-access \cite{aqeel2016security}. Solutions for privacy-preserving Edge Intelligence must therefore be effective in providing information security \cite{mann2022security}. There are 4 main privacy requirements during the Inference phase in MLaaS: the client may not learn 1) the model's architecture and 2) the model's trained parameters, while the party holding the model, typically the server, must not learn 3) the Inference input data nor 4) the Inference output. Here, we assume that standard system security methods (e.g., limiting the number of Inference requests) are in place to protect a fifth piece of potentially sensitive data, the training dataset (see model inversion attacks) \cite{2}.

General characteristics of Edge Intelligence are described extensively in \cite{yu2017survey, 210, xu2021Edge, 139}, providing criteria for privacy-preserving solutions applied in an Edge Intelligence context. Solutions should allow practical implementations in a commercial setting, and must provide  accurate and timely Inference. 

Edge intelligence setups may involve more than two parties during DNN Inference. For example, in a smart home sensor-actuator setup, computations offloaded to several servers (operated by different companies) may receive Inference input from some sources, while sending Inference output to other devices. Information should remain private, even if parties are secretly colluding. 

In an Edge setup, clients sending Inference inputs are often low-capacity devices with minimal compute capacity for data collection, temporary storage, and transmission tasks, while the model is held and evaluated by a nearby Edge server.

Servers are capable nowadays of receiving and sending more than a Gbps using LAN or other fast intranet networks. A potential communication bottleneck arises however, when network transmission is of type PAN (e.g., Bluetooth), WAN, MAN, or LPWAN, all common to Edge Computing setups.

\begin{table*}[htbp]
 \centering
     \caption{Summary of our assessment of Privacy-Preserving techniques for Edge Intelligence.}
\begin{tabular}{p{18mm}p{48mm}P{18mm}P{10mm}P{11mm}P{15mm}P{13mm}P{8mm}}
\hline
&&Fully\newline Homomorphic Encryption&Garbled Circuit&Secret Sharing&Model Splitting w/o noise&Model Splitting w/ noise&Secure Enclave\\
\hline
\multirow{9}{=}{Edge Intelligence\newline Requirements}
&fulfills the 4 privacy requirements&\cmark&\xmark&\cmark&\xmark&\xmark&\cmark\\
& can involve $>$2 parties&
\xmark&\xmark&\cmark&\cmark&\xmark&\xmark\\
&Inference accuracy&\cmark&\cmark&\cmark&\cmark&\xmark&\cmark\\
&low latency expected &\xmark&\xmark&\cmark&\cmark&\cmark&\cmark\\
&minimal compute capacity (client)&\xmark&\xmark&\cmark&\cmark&\cmark&\cmark\\
&limited compute capacity (server)&\xmark&\xmark&\cmark&\cmark&\cmark&\cmark\\
&limited communication&\xmark&\xmark&\xmark&\xmark&\cmark&\cmark\\
&high drop-out rate (client)&\cmark&\xmark&\xmark&\cmark&\xmark&\xmark\\
&hardware independence &\cmark&\cmark&\cmark&\cmark&\cmark&\xmark\\
\hline
\end{tabular}
\label{table1}
\end{table*}

Client drop-out occurs mostly to mobile devices such as smartphones, which can also easily turn off. In most cases, client dropout is inconsequential: even if the client is assigned chunks of Inference computations, DNN Inference can be paused until the client is within reach again. This criterion is however important for cases where device drop-out would disrupt task distribution (e.g., swarm intelligence with drones, and mobile computing).

Lastly, to make IoT objects available to consumers at affordable costs, they may lack state-of-the-art hardware specialized in DNN Inference. Additionally, when a device drops out, Edge Computing algorithms aim to dynamically re-assign tasks to the next available device regardless of hardware. Solutions should therefore be applicable to most types of hardware, without assuming that specialized hardware is available. 

\section{Assessment of Privacy-Preserving Techniques}
The main techniques that constitute the field of Privacy-preserving DNN Inference are reviewed in several comprehensive surveys \cite{2, 220, pulido2021privacy, ball2019garbled}. In this section, we assess the compatibility of each category of techniques with the requirements of Edge Intelligence (summarized in Table \ref{table1}). 

\textbf{Fully Homomorphic Encryption (FHE)}: a form of encryption \(E\) performed by the client to input data \(x\). \(E(x)\) is sent to the server, which returns after Inference, the cyphertext \(E(y)\), to the client. \(E(y)\) is the encrypted equivalent of the correct Inference output \(y\). Optionally, some parameters of the DNN can also be encrypted. FHE meets all four privacy requirements, and the Inference task itself can be completed in case of client drop-out. However, low-end client devices cannot carry the heavy cryptographic operations required. Despite promising recent advances \cite{258, reagen2021cheetah, lee2022privacy} since its inception \cite{135}, FHE schemes are still too computationally demanding for applications to Edge Intelligence. Additionally, FHE only supports additions and multiplications, and thus require additional processing for other non-linear operations (e.g., polynomial approximation of some activation functions). Despite this, little to no loss in accuracy has been reported, especially via re-training modified DNNs.

\textbf{Garbled Circuits}: 2-party protocol based on converting a neural network to a Boolean circuit made of AND, XOR, and XNOR gates, where each gate corresponds to an operation. The architecture of the DNN is known to both parties, but the input data and the weights of the DNN are kept secret. As a sub-protocol, oblivious transfer is used, a public-key cryptography-based scheme that enables one party to send one of two inputs to a second party so that the second party only learns one of the inputs and the first party does not learn which input the second party learned. Garbled Circuits support both linear and non-linear operations but are computationally costly (especially for AND gates \cite{KolesnikovS08}). Moreover, creation of the garbled circuit includes creation and permutation of a truth table per gate, to further encrypt it (e.g., using AES-based cryptography). As with FHE, this method is thus ill-adapted for low-end clients.

\textbf{Secret Sharing}:~\(n\)-party secure Multi-Party computing protocols in which each value involved in DNN Inference (i.e., input and model parameters) are  divided into~\(n\) shares, such that individual shares do not reveal anything about the secret values. Since originally introduced \cite{266}, several different secret sharing schemes have been proposed. Additive \cite{70, 281} and replicated \cite{274, wagh2021falcon} secret sharing seem especially appropriate for DNN Inference. For a complete Inference, the evaluation of the layers of the DNN may be performed in several ways, depending on various factors. In particular, different protocols can be used for addition, multiplication (e.g., masking inputs with Beaver Triplets), and non-linear operations (e.g., polynomial approximation, garbled circuits). The protocols also depend on the number of parties involved, as well as whether colluding is accounted for or not. Servers can also send a client shares to compute. Secret Sharing is not encryption-based and therefore relatively cheap to add to DNN Inference tasks. DNN Inference however fails if devices holding information on how shares are created (e.g., random number generator) drop out. Secret Sharing is still a communication-intensive privacy-preserving method, necessitating a high number of communication rounds.

\textbf{Model Splitting without noise}: partitioning a DNN so that each party receives unprocessed chunks of calculations, including raw weights and inputs. Accuracy is therefore preserved. The higher the number of devices recruited, the higher the privacy as well as speed (with possibility of parallel computing), as no party may reconstruct the neural network, nor infer the training nor input data from the parts it receives \cite{285}. This method requires the client to perform the initial and last computations, but does not need a powerful server. In case of colluding, model architecture and parameter privacy are largely lost.

 \begin{table*}[htbp]
 \label{table:table2}
 \centering
 \caption{Summary of the impact of Model Compression techniques on Secret Sharing for DNN Inference}
\begin{tabular}{p{40mm}P{25mm}P{30mm}P{20mm}P{30mm}}
\hline
&Less operations\newline in total&Less non-linear operations&Reduced message sizes&Less  communication rounds\\
\hline
Quantizing&yes&yes&yes&no\\
Pruning&yes&not purposefully&yes&not purposefully\\
Knowledge Distillation&yes&yes&no&yes\\
Low-rank approximation&yes&yes&no&yes\\
\hline
\end{tabular}
 \label{table2}
 \end{table*}
 
\textbf{Model Splitting with noise}: noise is added by the client to the input data, intermediary results, and/or weights when the client receives partial computations from a DNN. The noise added must fulfill the requirements for Differential Privacy, which mathematically guarantee that data is obfuscated sufficiently to conceal individual records (e.g., a person’s identity) it may contain. This is necessary because raw input data can be reconstructed from intermediary results after even~6 layers \cite{he2020attacking}. There is a privacy/accuracy trade-off based on the amount of noise added. The client is responsible for noising and de-noising, which can be computationally expensive depending on the scheme used (e.g., auto-encoders and decoders for obfuscation).

\textbf{Secure Enclaves}: are dedicated portions of memory which are designated by the CPU as inaccessible from the operating system nor any other application, and within which data can be secretly processed and encrypted/decrypted if necessary. A popular example of Secure Enclaves is Intel's SGX \cite{sgxtz}. For DNN Inference, two parties may send an encrypted model and input data, respectively, which can then be decrypted and processed within Secure Enclaves, finally returning the Inference output to the appropriate party, thus providing full privacy. Secure Enclaves are however costly and more memory limited than traditional hardware.

This assessment indicates particular suitability of Secret Sharing for Edge Intelligence, as it meets the most criteria (7 out of 9), especially meeting all information privacy and performance-related requirements. Therefore, we dedicate the rest of the paper to discuss secret sharing and its applicability for Edge Intelligence.

\section{Implications of State-of-the-Art Performance}
Recent solutions for Secret Sharing in DNN Inference \cite{220}, not only for Edge Computing, all still use outdated Convolutional Neural Networks (CNNs) as evaluative benchmarks (e.g., AlexNet \cite{krizhevsky2017imagenet} trained on MNIST \cite{deng2012mnist}). The performance of these CNNs is humble compared to that of Transformers (e.g., answer generation from multi-modal inputs \cite{li2019visualbert}), a state-of-the-art category of DNNs now ubiquitous in the field of AI \cite{zaidi2022survey}. 

The question then arises: would the solutions, as they are, be applicable for fast Edge Intelligence in the context of a real and state-of-the-art MLaaS task? The answer is probably `no'.

Primarily, larger DNNs have higher complexity (i.e., more parameters to compute, more nodes in layers due to larger inputs), leading to an increase in the number of secret shares to produce and re-combine. The amount of non-linear operations during Secret Sharing, while manageable on smaller DNNs, can become problematic as it increases, which may necessitate more Garbled Circuits and/or Beaver Triplets, and both methods are especially intensive in 2-party settings, despite possibilities of some offline processing \cite{mann2022towards}. 

Furthermore, state-of-the-art DNNs have a higher diversity in the types of layers, (e.g., Self-Attention, Recurrent) than benchmark CNNs do which current Secret Sharing schemes are not yet designed to handle \cite{mann2022towards}. New types of layers (e.g., Self-Attention) have more data processing, such as parallel encodings of subsets of inputs (e.g., a single word), as well as more operations to perform per layer than classic layers (e.g., ReLu, pooling), requiring new protocols other than current ones which consider a layer as a unit only taking in simultaneous inputs \cite{220}. 

\section{Model Compression and Secret Sharing}
A first step towards bringing large Transformers to Secret Sharing particularly for Edge Intelligence, is to tackle the computation and communication bottleneck. Three categories of solutions exist: 1) Hardware Acceleration \cite{87}, consisting of a set of instructions to parallelize computational tasks into specialized hardware components (e.g., Neural network Processing Units -- NPUs \cite{xpu}) -- similarly to Secure Enclaves, they may be too expensive to integrate in commercial objects; 2) Software Orchestration \cite{softacc,aghapour2022}, consisting of developing data pipelines or algorithms to optimize resource management to reduce latency of DNN Inference -- it is assumed to be applied to some extent in any distributed system, and cannot reduce the rounds of communication required for Secret Sharing; and 3) Model Compression \cite{xu2021Edge, compresssurvey}, reducing the amount and complexity of the computations -- typically requiring re-training to retain accuracy.

Model Compression techniques, namely quantization, pruning \cite{liang2021pruning}, knowledge distillation \cite{distil}, and low-rank approximation \cite{idelbayev2020low}, while increasingly customary in Edge Intelligence, have yet to be compared in the context of Secret Sharing. Table \ref{table2} provides a qualitative comparison. 

Quantization reduces the byte-size of each value, and consequently the size of the secret shares communicated between parties, but leaves the total number of communication rounds largely unaffected. Particularly, solutions with binary or ternary quantization to weights, and a limited fixed-point size to activation functions, preserve the granularity of input data while significantly reducing communication \cite{liang2021pruning}. Quantization can be combined with other model compression techniques as well.

Pruning removes inconsequential computations in a DNN. It offers no guarantees of effectiveness in addressing computational complexity specific to Secret Sharing. In the best cases, however, Pruning may remove a significant number of connections between the nodes of a DNN, thus reducing computation and communication.

Knowledge Distillation (i.e., training a smaller network off of a larger one) and Low-rank Approximation (i.e., reducing the dimensionality of each layer via matrix decomposition) are more promising candidates for secret sharing. Firstly, they remove extra features from the input data sooner, thus reducing the amount of input to propagate throughout the network. Secondly, they both reduce the amount of computations systematically throughout the DNN (i.e., most layers are reduced in size). Consequently, less rounds of Secret Sharing communication are necessary per layer. Knowledge Distillation also reduces the number of layers \cite{zaidi2022survey}. Lastly, both Knowledge Distillation and low-rank approximation are actively researched and have recently been successfully applied to state-of-the-art transformers (e.g., BERT \cite{devlin2018bert} became DistilBERT \cite{sanh2019distilbert} and Ladabert \cite{mao2020ladabert}). The accuracy of compressed versions of those model is also improving \cite{zaidi2022survey}. 
 
\section{Conclusion}
Secret Sharing was deemed the most promising privacy-preserving technique given Edge Intelligence characteristics, but is not yet applicable to state-of-the-art Deep Neural Networks. Future research should address the new types of DNN layers and computations that current Secret Sharing schemes do not yet account for, while optimizing performance for MLaaS in Edge Computing. We put forward, pending experiments, Knowledge Distillation and Low-Rank Approximation as promising means to further accommodate new Secret Sharing protocols, for practical Edge Intelligence.


\bibliography{chabal23}

\end{document}